\documentclass[fleqn,usenatbib]{mnras}

\usepackage{newtxtext,newtxmath}

\usepackage[T1]{fontenc}
\usepackage{ae,aecompl}

\usepackage{graphicx}	
\usepackage{amsmath}	
\usepackage{amssymb}	

\usepackage{pdflscape}
\usepackage{upgreek}

\title{Particle acceleration in the Herbig-Haro objects HH 80 and HH 81}

\author[A. Rodr\'iguez-Kamenetzky]{Adriana Rodr\'iguez-Kamenetzky$^{1}$\thanks{E-mail: adriana.rodriguez@unc.edu.ar},
Carlos Carrasco-Gonz\'alez$^{2}$, Omaira Gonz\'alez-Mart\'in$^{2}$,
\newauthor Anabella T. Araudo$^{3}$, Luis Felipe Rodr\'iguez$^{2}$, Sarita Vig$^{4}$, Peter Hofner$^{5}$
\\
  $^{1}${Instituto de Astronom\'ia Te\'orica y Experimental, (IATE-UNC), X5000BGR C\'ordoba, Argentina}\\
  $^{2}${Instituto de Radioastronom\'ia y Astrof\'isica (IRyA-UNAM), 58089 Morelia, M\'exico}\\
  $^{3}${Astronomical Institute of the Czech Academy of Sciences, Bocni II 1401, Prague, CZ-14100 Czech Republic}\\
  $^{4}${Dept. of Earth and Space Science, Indian Institute of Space science and Technology, Thiruvananthapuram, 695 547, India}\\
  $^{5}$Physics Department, New Mexico Tech, 801 Leroy Place, Socorro, NM 87801, USA
}

\pubyear{2018}

\hypersetup{draft}
\begin{document}
\label{firstpage}
\pagerange{\pageref{firstpage}--\pageref{lastpage}}
\maketitle

\begin{abstract}

We present an analysis of radio (Karl G. Jansky Very Large Array (VLA)), optical (HST), and X-ray ({\it Chandra} and {\it XMM-Newton}) observations and archival data of the Herbig-Haro objects HH~80 and HH~81 in the context of jet-cloud interactions. Our radio images are the highest angular resolution to date of these objects, allowing to spatially resolve the knots and compare the regions emitting in the different spectral ranges. We found that soft X-ray thermal emission is located ahead of the non-thermal radio peak. This result is consistent with a radiative forward shock that heats the shocked gas up to 10$^6$K, and an adiabatic reverse shock able to accelerate particles and produce synchrotron radiation detectable at radio frequencies. These high angular resolution radio images also reveal a bow shock structure in the case of HH~80N, being the first time this morphology is detected in a Herbig-Haro object at these frequencies.

\end{abstract}



\section{Introduction}
 
 The earliest stages of star formation are characterized by strong accretion at the same time that the protostar drives powerful collimated winds (jets). One of the clearest manifestations of ejected material from protostars are the Herbig-Haro objects (HH), observed at parsec scales. These objects trace shocks produced by the interaction of a supersonic jet with the environment. Since their discovery (\citealt{Herbig1951} and \citealt{haro1952}), HH objects have been studied in different spectral ranges, e.g. optical, infrared, radio, and recently, also X-ray emission has been reported in a few of these (\citealt{pravdo2001}; \citealt{favata2002}; \citealt{bally2003}; \citealt{pravdo2004}; \citealt{tsujimoto2004}).

 Objects HH~80 and HH~81 (HH~80/81 henceforth) were discovered by \citealt{rei&gra1988} at the edge of the L~291 molecular cloud, that constitutes a place of recent star formation located at a distance of 1.7~kpc. The source IRAS~18162-2048 is associated with a massive B-type protostar surrounded by an accretion disk (e.g. \citealt{carrasco2012}, \citealt{girart2018}) and seems to be the powering source of HH80/81. HH 80 and HH 81 were the first HH objects discovered at radio frequencies (\citealt{rod&reip1989}), with a flux density of a few mJy. Later, their northern counterpart was identified at radio frequencies by \cite{marti1993} as HH~80 North (HH~80N). This deeply embedded object is the first HH recognized as such being only detected at radio frequencies; its HH nature  was later corroborated by photochemical effects detected in a molecular gas condensation and dust located in front of this object (\citealt{girart1994}, \citealt{girart1998}; \citealt{masque2009}).
 
 The protostar-disk system drives a very powerful jet, which shows radio emission up to 3~pc (e.g. \citealt{marti1993}). The HH~80-81 jet represents the most clear case of a protostellar jet with presence of non-thermal radio emission. Near its driving source (up to $\sim$0.5~pc), this emission shows negative spectral indices (\citealt{marti1993}; \citealt{arrk2017}). Polarized radio emission was also detected in this region of the jet, thus confirming its synchrotron nature (\citealt{carrasco2010Sci}). These previous works show that the HH~80-81 jet is powerful enough to accelerate particles at strong shocks. Despite being located at large distances from the protostar, HH objects also seem to present a non-thermal component of radio emission. This can be inferred from the detection of very negative spectral indices ($\alpha=-$0.3 a $-$0.7; e.g., \citealt{marti1993}, \citealt{vig2018}) measured in low angular resolution observations.
 
  Kinematic studies of HH~80, HH~81, and HH~80N (hereafter HH~80/81/80N), reveal proper motions of $\sim$200-400~km~s$^{-1}$ (\citealt{masque2015}, \citealt{HRR1998}), while the velocity of the jet material is about $1000~{\rm km~s}^{-1}$ (\citealt{marti1995}). This implies that the fast jet brakes during its interaction with the molecular cloud, giving rise to HH objects. According to theoretical models, particle acceleration can take place in this scenario (e.g., \citealt{araudo2007}, \citealt{bosch2010}, \citealt{romero2010}), which could explain the presence of synchrotron emission in these objects. These models predict emission in a wide range of wavelengths, ranging from radio to $\upgamma$-rays. However, these objects are usually embedded in molecular clouds, which makes their detection, at high energies, difficult due to extinction by foreground material. In the last years, both theoretical and observational studies have been carried out to investigate particle acceleration in protostellar jets (e.g., \citealt{arrk2016}, \citealt{arrk2017}, \citealt{ains2014}, \citealt{padovani2015}, \citealt{padovani2016}), nevertheless, it still remains unclear how particles accelerate in these systems. In this sense, Herbig-Haro objects HH~80 and HH~81 constitute an excellent case to study the jet-cloud interaction, since besides being detected at optical wavelengths, they are also detected at radio continuum and X-rays.
  
 Here we analyze radio (VLA\footnote{Very Large Array, of the National Radio Astronomy Observatory (NRAO). The NRAO is a facility of the National Science Foundation operated under cooperative agreement by Associated Universities, Inc.}), optical (HST), and X-ray ({\it Chandra} and {\it XMM-Newton}) emission, associated with these objects. Our high angular resolution radio images allows us to study their morphology and also compare the emitting regions in the different spectral ranges we worked with. This, along with a spectral analysis of radio and x-ray emission lead us to propose a scenario for the jet-cloud interaction.

\section{Observations}
\subsection{Radio continuum}

  Here, we analyze C-band data obtained with the VLA in 2009 and 2012. Data from 2009, previously presented by \citealt{carrasco2010Sci}, was obtained at C configuration with a 100~MHz bandwidth (see \citealt{carrasco2010Sci} for a full description of the calibration process). Data from 2012, reported in \citealt{arrk2017}, were obtained with the VLA in B configuration, with a bandwidth of 1~GHz (L-band) and 2~GHz (S and C bands) (a complete description of the data and calibration process can be found in \citealt{arrk2017}).

For this data set we obtained different images. We have built a 2~GHz bandwidth image centered at 5.5~GHz using {\it briggs} weighting (\citealt{briggs1995}) with {\it robust}\footnote{The {\it robust} parameter ranges from $-$2 (uniform weighting) to $+$2 (natural weighting)}=1.5. To study the morphology of these objects we combined 5.5~GHz data taken in C and B configurations (previously described). By combining these observations, it is possible to achieve images sensitive to both extended and compact structures associated with HH objects. Different weightings were used, taking the parameter \emph{robust} equal to -1 and 0. 

The values of angular resolution and sensitivity for the images we obtained are $\sim$ 1$''$-2$''$ and 15-18$\upmu$Jy/beam, respectively. On the other hand,  since the largest detectable angular scales are 30$''$ (B-configuration data) and 240$''$ (C/B-configuration data), the extended emission of HH objects ($\sim4''$) is well detected. Detailed information on the parameters of the images presented in this work are summarized in Table \ref{tbl-images_param}.

\subsection{X-rays}

Objects HH~80/81 were observed with the \emph{Chandra} satellite in 2006 (obsid 6405) by \cite{pravdo2009}, using the ``Advanced CCD Imaging Spectrometer'' (ACIS) (\citealt{garmire2003}). The data was downloaded from HEASARC\footnote{https://heasarc.gsfc.nasa.gov} archives, and processed with the software ``Chandra Interactive Analysis of Observations'' (CIAO) (\citealt{frus2006}), version 4.6, with upgraded calibrations (CALDB version 4.6.3). Images in different bands were made using the {\sc dmcopy} and {\sc csmooth} tools of CIAO: 0.3-1.2 keV (soft), 4.5-10.0 keV (hard), and 0.3-10.0 keV (total); the pixel scale was 0.125 arcsec. To favor the detection of extended structures we used a Gaussian smoothing Kernel whose FWHM was 3-4 pixels, depending on the local S/N ratio.

We also analyzed observations of HH~80 and HH~81 performed with \emph{XMM-Newton} in 2003 (obsid 0149610101, P.I. Steven Pravdo) with a total exposure time of 47 ksec, available in the HEASARC archive. These observations were made using the EPIC-pn detector only, which offers the most sensitive spectra; the MOS detectors were not used to avoid cross-calibration problems between instruments. Spectral extraction and high-period background filtering process were performed with the software SAS version 16.1.0 (\citealt{gabriel2004}), especially developed for \emph{XMM}-Newton. High-period background events can be due to solar flares and/or cosmic rays, and must be removed from the observations in order to not affect the results. With this purpose we extracted the light curve from the background to detect periods of time where the background is high.

\subsection{Optical}
In this paper we compare radio and X-ray emission with optical images taken with the Hubble Space Telescope. These images were obtained and published by \citealt{HRR1998}, and correspond to three atomic line filters: H$\alpha$+[NII], [SII], and [OIII]. In the case of HH objects, these transitions are commonly excited by collisions, being H$\alpha$+[NII] and [SII] usually detected with relatively intense emission. Nevertheless, the high excitation line [OIII] is only detected when strong shocks take place, implying shock velocities of $\sim$ 700~$\rm{km~s^{-1}}$ (\citealt{HRR1998}). These images are the highest quality and angular resolution image obtained for HH~80/81 in the optical range, and are only used with the purpose of comparing the morphology of these objects with their emission at radio and X-ray frequencies.

\section{Results and Analysis}
 
We analyze and discuss the optical, X-ray, and radio emission associated with Herbig-Haro objects HH~80/81, and their counterpart, HH~80N, detected at centimeter wavelengths only.

\subsection{Radio continuum}
 
In order to illustrate the position of the HH~80/81 objects with respect to the driving source of the HH~80-81 jet, we show in Figure \ref{structure} (panel a) a radio continuum image at 5 GHz previously reported by \citealt{carrasco2010Sci}, where the full extension of the jet can be seen ($\sim$ 7.5~pc, \citealt{HRR1998}) with relatively low angular resolution (13''$\times$8'', PA=2$^{\circ}$). In this image we label the central source associated with the protostar (IRAS 18162-2048) and Herbig-Haro objects HH~80/81/80N. Also, in the innermost region of the jet, the structure of two lobes emerging from the central source in opposite directions (NE-SW) is clearly identified with a total extension of $\simeq$1~pc. Panels b, c, and d, show the HH objects with higher angular resolution (of about $2''\times1''$ for HH~80N, and $3''\times2''$ for HH~81 and HH~80).

 \subsubsection{Morphology}

\begin{figure*}
   \centering
    \includegraphics[scale=0.45]{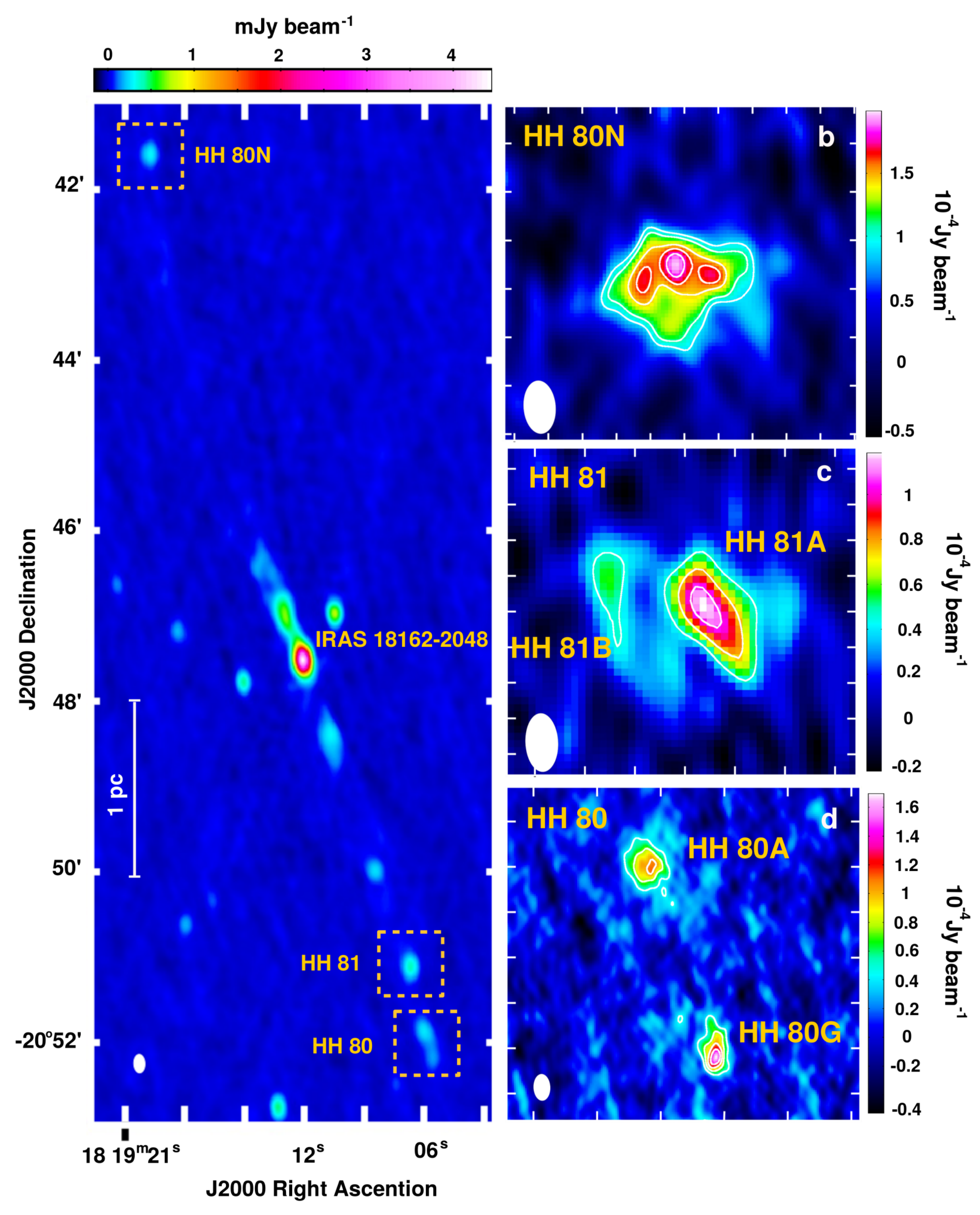}
    \caption{Radio continuum images at 5.5~cm of the HH 80-81 jet. The beam size is shown in the lower left corner of each panel. (a) Image obtained with the JVLA radio interferometer in C configuration with a synthesized beam of 13$''\times8''$ and PA 2$^{\circ}$, previously reported by \citealt{carrasco2010Sci}. In this panel we present the full extent of the HH~80-81 jet; Herbig-Haro objects and the radio source associated with the driving source of the jet (IRAS 18162-2048) are labeled. (b) {\bf HH~80N:} contours correspond to intensity levels of 5, 6, 8, 9, and 10 times the rms (18~$\upmu$Jy~beam$^{-1}$) in the image obtained in 2012 (B configuration) using Briggs weighting (robust= 1.5, beam $2\farcs19\times 1\farcs32$;~PA 4\fdg5). (c) {\bf HH~81:} Image that combines 2009 (C conf.) and 2012 (B conf.) data, using Briggs weighting (robust= -1, beam $=1\farcs64\times 0\farcs91$;~PA $2\fdg78$); contours correspond to intensity levels of 3, 5, and 7 times the rms, 15$\upmu$Jy~beam$^{-1}$. (d) {\bf HH~80:} Image that combines 2009 (C conf.) and 2012 (B conf.) data, using briggs weighting (robust= 0, beam $=2\farcs92\times 1\farcs83$;~PA $2\fdg43$); contours correspond to intensity levels of 3, 5, 7, and 9 times the rms, 15~$\upmu$Jy~beam$^{-1}$. Knots HH~81A, HH~81B, HH~80A, and HH~80G identified by \citealt{HRR1998} at optical wavelengths are labeled in panels (c) and (d).}
   \label{structure}
   \end{figure*}

To study the morphology of HH~80/81/80N at centimeter wavelengths, we analyze images that combine JVLA data corresponding to different configurations (see Table \ref{tbl-images_param}). These images allow the detection of compact emission regions, without losing information on extended structures, and represent the highest angular resolution radio images obtained to date for these objects. Thereby, for the first time, it is possible to study the radio morphology of HH objects in this system, and its relation with components dominated by emission in different spectral ranges. In Figure \ref{structure} we show the objects HH~80N, HH~81, and HH~80 in panels b, c, and d, respectively, whereby
the knots identified at optical wavelengths by \citealt{HRR1998} are labeled. A comparison of the locations of radio continuum emission and optical emission lines reveals that the radio emission is tracing shocks produced by the interaction of material ejected by the protostar with the ambient medium. With the angular resolution achieved, we can see that both HH~80 (A and G) and HH~81 (A) are elongated in the jet direction, having its maximum emission at the apex, consistent with what is expected to be observed in a marginally resolved bow-shock. However, to be able to observe the working surface morphology, higher angular resolution observations are needed. In this sense, the object HH 80N constitutes an interesting case, since our high angular resolution images allow for the first time to resolve its morphology, showing details that can not be seen in HH 80/81. Due to the high extinction produced by the molecular cloud, this object is only detected at radio wavelengths, and reveals a complex structure, e.g., flattened at the frontal region of the shock, with presence of high intensity spots and a weak emission tail elongated towards the protostar (see Figure \ref{structure}b).

\subsubsection{Time variability}
Regarding time variability of the radio emission, most of the radio jets that have been monitored show no variability above a 10-20\% level. There are, however, cases where variability has been detected on timescales of years (\citealt{anglada2018}). The study of the time variability of HH objects is even more limited than that of radio jets. For example, in the case of HH~1 and HH~2 variations in the optical and radio emission of up to a factor of two have been found over timescales of 20 years (\citealt{LF2018}). Assuming this variation was smooth, it implies that over a timescale of 3 years (as in our case) we could expect variations of order 15\%. We do not expect similar variations to affect significantly our conclusions.

 \subsection{X-ray emission}
 To study the X-ray emission detected in the region of HH~80/81 we analyze archival data from Chandra and XMM-Newton missions in three spectral ranges, i.e., hard: 4.5-10.0~keV; soft: 0.3-.2~keV; total: 0.3-10.0~keV. The high resolution images obtained with Chandra allow us to identify soft and hard emission regions associated with these objects, while high sensitivity data obtained with XMM-Newton, allows us to specify the nature of the dominant emission in the range 0.3-10.0~keV through a spectral analysis. 
 
  The spectral fit was performed with the X-ray Spectral Fitting Package {\it XSPEC}. We extracted the spectrum of each source in circular regions in the range 0.3-10.0 keV, considering events with patterns 0-4, and using the {\sc dmextract} function of SAS. These circular apertures were chosen to be the smallest that include the extended emission of the HH objects, and preserve a good S/N ratio, i.e., 30$''$ (HH~80) and 17$''$ (HH~81) diameter. The spectral fit uses the sky background information simultaneously, selected from a region located in the vicinity of the sources. Finally, to ensure a reliable statistic $\rm{\chi^{2}}$, each spectrum was sampled to have at least 15 events per bin in the spectrum extracted from the sky background. This procedure was performed with the function {\sc grppha} of FTOOLS. In Figure \ref{xray} we show the X-ray emission detected in the region of HH~80 (bottom panels) and HH~81 (upper panels) with Chandra (contours) and XMM-Newton (color scale), in the three spectral ranges mentioned above. Magenta circles indicate the regions where the XMM-Newton spectra were extracted.
  
  In both cases, the spectral analysis shows that the emission at low energies ($\lesssim$1~keV) can be described by a thermal emission model for hot diffuse gas\footnote{The model used was {\it mekal}, based on Mewe and Kaastra calculations (\citealt{mewe1985}; \citealt{mewe1986}; \citealt{lied1995}).}, however this model seems to underestimate the emission at energies higher than $\sim 1$~keV. We also tried to fit the spectrum by combining two thermal components, but this implies the presence of diffuse gas at excessively high temperatures ($\sim 100\times10^{6}$~K). Such high temperatures require shock velocities about 2000-3000 km~s$^{-1}$, and therefore, very high jet velocities ($\rm v_{jet}>v_{shock}$, since the knots are braking), which does not represent a realistic scenario in the case of HH objects. Therefore, the spectrum in the range 0.3-10.0~keV cannot be described by a thermal emission model only. The best fit ($\chi^{2}$ = 30.1 with 28 degrees of freedom) was obtained by the addition of a thermal emission component and a non-thermal component, modeled by a power law of the form $A(E) = K E^{-\upgamma}$ (being $\upgamma$ the dimensionless photon index in the power law, and $K$ the normalization factor [photons/keV/cm$^2$/s at 1 keV]). Both components are affected by a photo-electric absorption model $M(E)=\exp[-{\rm N}_H\sigma(E)]$, being $\sigma(E)$ the photo-electric cross section (\citealt{morrmcc1983}), and $N_{H}$ the hydrogen column density (in units of $10^{22}$ atoms~cm$^{-2}$). By knowing $N_{H}$, the visual extinction ($A_{V}$) can be estimated with the expression $N_{H}~(\rm cm^{-2})=(2.21\pm0.09)\times 10^{21}A_{V}~(\rm mag)$ (\citealt{guvoz2009}). Thus, we calculate a visual extinction of $A_{V}= 2.79$~mag and $A_{V}= 3.37$~mag for HH~80 and HH~81, respectively. These values are near to those obtained by \cite{HRR1998} in the optical range ($A_{V}= 2.33$~mag). The spectral fit is shown in Figure \ref{spec_hh}, and the derived parameters ($N_{H}$, kT, and $\upgamma$) are listed in Table \ref{tbl-spec}.
 
 In the following sections we analyze the soft and hard emission detected in the region of the HH~80/81 objects. 
 
\subsubsection{Soft X-rays}

 In the images obtained with XMM-Newton data (Figure \ref{xray}, color scale) we can clearly see that in both cases the dominant radiation associated with HH objects is detected in the soft band. In the same energy range, Chandra images (contours) reveal that this emission comes from a more compact region, whose morphology is elongated in the direction perpendicular to the jet axis. The spectral analysis shows that soft X-ray emission is consistent with thermal emission from diffuse gas with temperatures of the order $\sim10^{6}~K$ (Table \ref{tbl-spec}), that could be heated by compression in a radiative shock produced by the interaction of the jet with the molecular cloud, as we discuss in Section~\ref{termination-shocks}. Knowing the temperature of the gas and using the expression presented by \cite{raga2002}, it is possible to estimate the propagation velocity of the shock ($V_{\rm shock}=(T/15)^{1/2}~{\rm km~s^{-1}}$, being T the temperature of the shocked gas, in K). Thus, for HH~80 and HH~81 we obtain a velocity of $\sim 300$~km~s$^{-1}$, in agreement, within errors, with values derived for the brightest knots identified at optical wavelengths by \cite{HRR1998}, and radio continuum (6~cm) by \cite{masque2015} (See Section \ref{multi}).

\subsubsection{Hard X-rays} 
 
\begin{landscape}
   \begin{figure*}
   \centering
   \includegraphics[scale=0.5]{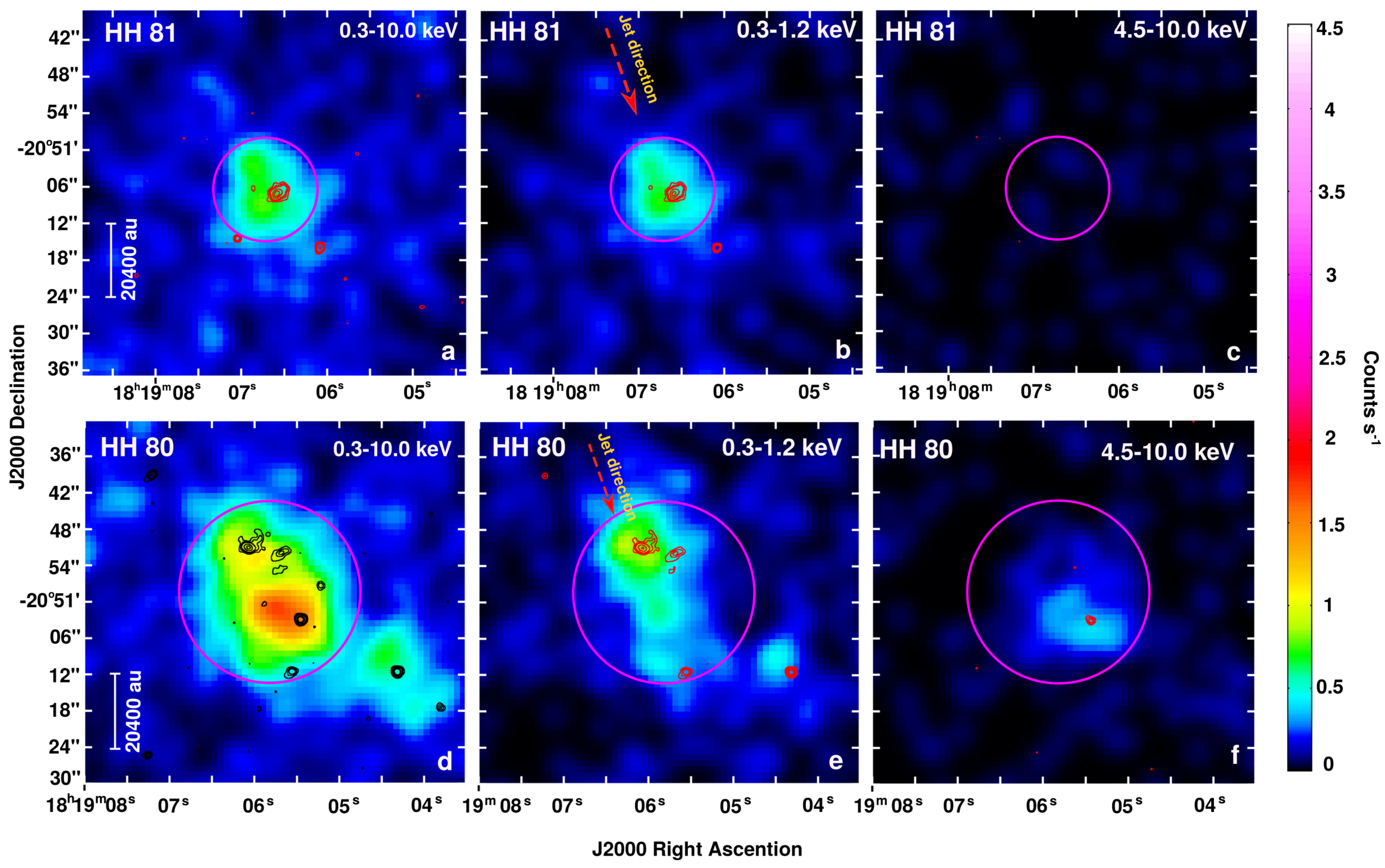}
   \caption{{\bf HH~80 and HH~81}: Superposition of X-ray emission in different energy ranges, detected by the missions XMM-Newton (color scale) and Chandra (contours). Upper and lower panels correspond to HH~81 and HH~80, respectively. Left panels show the total X-ray emission (0.3-10.0~keV). Soft (0.3-1.2~keV) and hard (4.5-10.0~keV) X-ray emission are shown in central and right panels, respectively. In the figure we also indicate the jet direction, i.e. the direction towards the driving source of the jet (dashed line), and the area where XMM-Newton spectra were extracted (circle).}
   \label{xray}
   \end{figure*} 
   \end{landscape}

\begin{figure*}
   \includegraphics[width=1\textwidth, scale=0.3]{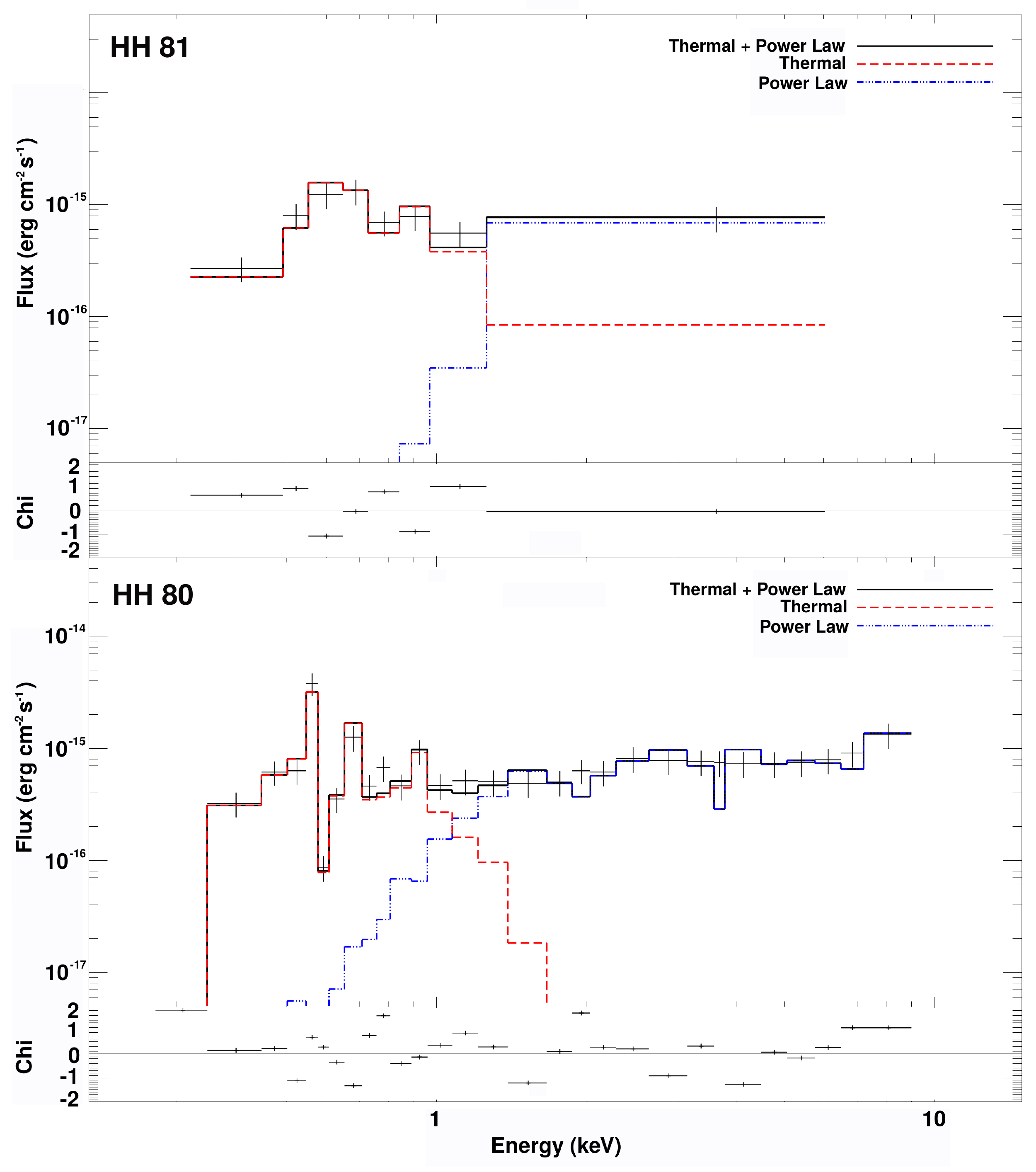}
   \caption{XMM-Newton spectra in the range 0.3-10.0~keV: HH~80 (upper panel) and HH~81 (lower panel). Observed data are shown with black crosses, whose length indicates the measurement errors. Dashed (red) and dot-dashed (blue) lines correspond to the thermal and non-thermal fits, respectively, while the solid line shows the addition of both components.}
   \label{spec_hh}
   \end{figure*}

 Hard X-ray emission is clearly detected in the XMM Newton images (Figure \ref{xray}, color scale) and spectrum (Figure \ref{spec_hh}, lower panel) of HH 80. A similar detection in the image is not clear in the case of HH 81, but the spectrum of this source shows emission in the hard X-ray spectrum (Figure \ref{spec_hh}, upper panel). In both cases, the emission in the hard band can only be interpreted in therms of a non-thermal emission component.
 
 In the case of HH~80, Chandra images (contours in Figure \ref{xray}, panel f) reveal the presence of a very compact source. This source has previously been interpreted by \cite{lopezsantiago2013} as synchrotron radiation of particles accelerated in the frontal shock. However, the analysis we present in this work suggests that it could be the case of a background source, since this compact source does not seem to be associated with emission detected in any other spectral bands (see Section \ref{multi}). In order to investigate the nature of this compact hard source detected by CHANDRA, we obtained total fluxes by integrating the spectrum in the different bands in both, CHANDRA and XMM Newton, and we compared them. From this, we estimated upper limits to the contribution of this source at the different energy ranges. For instance, the contribution of this source to the whole spectrum (0.3-10 keV range) is less than 10\%, and the contribution to the hard part of the spectrum (4.5-10 keV) is less than 20\%. The contribution to the soft band (0.3-2 keV) where the thermal X-ray emission is detected, is less than 5\%. Then, from this we conclude that the contribution of this source to the thermal spectrum is negligible. We also conclude that the compact hard X-ray source detected by CHANDRA is not able to explain all the emission detected in the hard band with XMM Newton. On the other hand, the spectrum of HH~81 also reveals the existence of a hard non-thermal component that, in this case, cannot be detected in the images taken with Chandra nor XMM-Newton. 

 These evidences suggest that X-ray non-thermal emission is actually arising from a diffuse and faint component present in both HH~80/81. This emission becomes detectable only in the spectrum integrated over all the region surrounding the HH objects. Moreover, the photon spectral index of the power law is, in both cases, $-$0.9 (see Table \ref{tbl-spec}). This gives a nearly flat energy spectral index of +0.1, that definitely rules out extension of radio synchrotron to X-ray regime. One possible interpretation is that this emission could be due to inverse Compton scattering (IC) of thermal infrared from the ambient cloud, produced by relativistic particles accelerated in the reverse shock, or present in the molecular cloud (e.g. \citealt{vig2018}). However, to identify the processes that can produce this radiation, a deeper and detailed study is required, using higher sensitivity X-ray images.

\subsection{Multifrequency analysis: optical, radio, and X-rays}\label{multi}

 Theoretical models of shocks produced by the interaction of protostellar jets with the ambient medium, predict emission over a wide range of wavelengths, from radio to gamma-rays (e.g., \citealt{araudo2007}; \citealt{bosch2010}; \citealt{romero2010}). Nevertheless, the absorption produced by the molecular cloud hinders the detection of high energy radiation in these shocks, and therefore, only in a few cases has it been possible to study the X-ray emission associated with them.

\begin{figure*}
    \centering
    \includegraphics[scale=0.65]{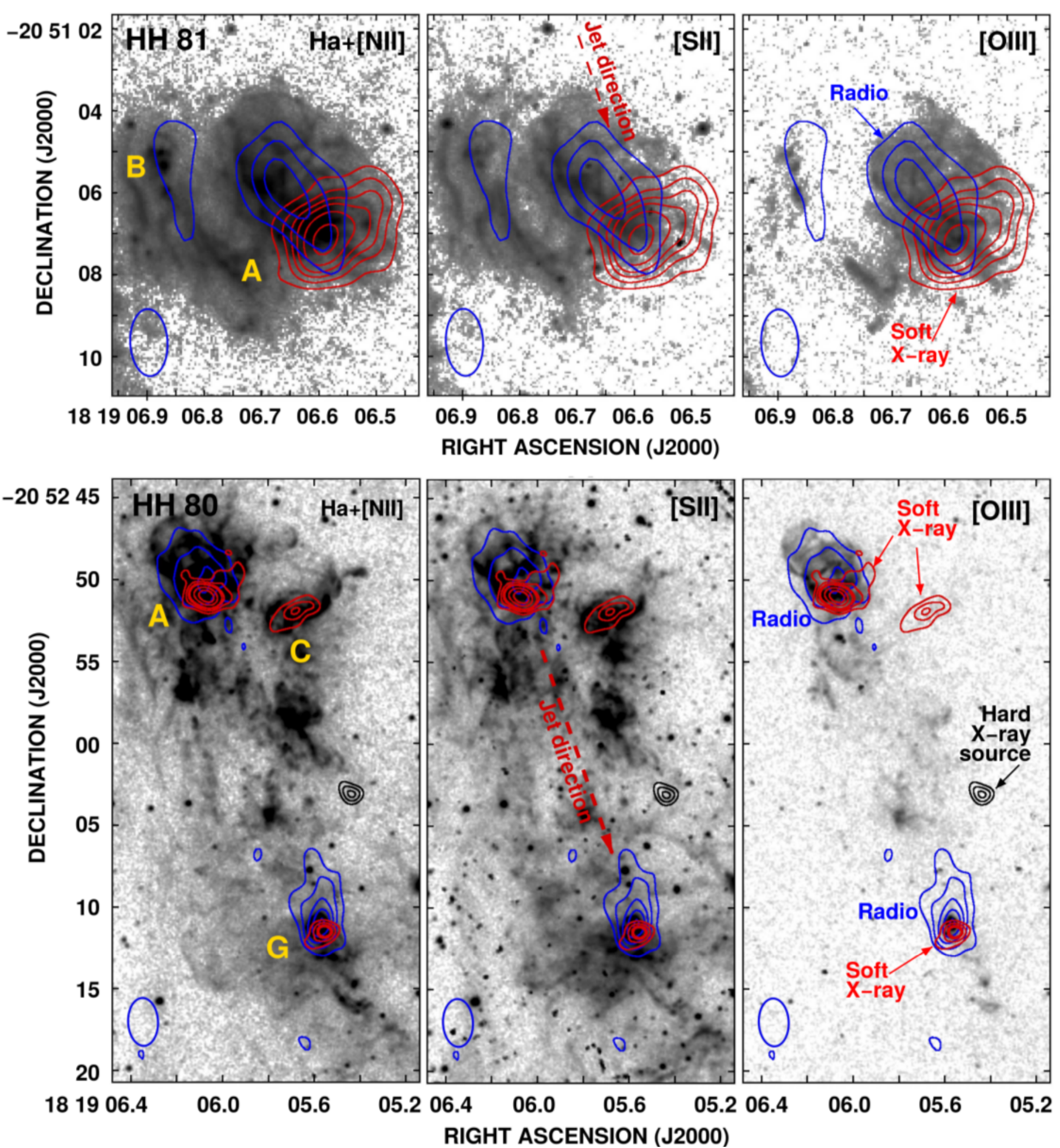}
    \caption{{\bf HH 80 and 81}: Image composed of the superposition of data in the optical range (gray scale), radio (blue contours), and X-rays (red and black contours). In gray scale we show optical images obtained with the HST in three filters (H$\alpha$+[NII], [SII], and [OII]). In the central panels we indicate the jet direction, i.e., the direction from the central source IRAS (dashed arrow), while in the left panels, we have labeled the knots identified by \citealt{HRR1998} at optical wavelengths: A, B, C, and G. {\bf HH 81} (upper panels): Radio continuum emission at 5.5~GHz (briggs weighting, robust=-1, beam size $=1\farcs64\times0\farcs91$;~PA $2\fdg78$); blue contours are intensity levels corresponding to 3, 5, and 7 times 15 $\upmu$Jy~beam$^{-1}$. Red contours correspond to X-ray emission detected by Chandra (soft band: 0.3-1.2 KeV). {\bf HH 80} (lower panels): Radio continuum emission at 5~GHz (briggs weighting, robust=0, beam size$=2\farcs92\times1\farcs83$;~PA $2\fdg43$); blue contours are intensity levels corresponding to 3, 5, 7, 9, and 11 times 15 $\upmu$Jy~beam$^{-1}$). X-ray emission detected by Chandra is shown in red (soft band, in the range 0.3-1.2 KeV) and black (hard band, in the range 4.5-10.0 keV) contours.}
    \label{orx}
    \end{figure*}

 In Figure \ref{orx} we present images of HH~80/81 showing X-rays detected with Chandra (red and black contours) and 5.5 GHz radio continuum (blue contours) over optical images obtained with the HST in three filters (H$\alpha$+[NII], [SII], and [OIII]). These filters correspond to the emission of species with different excitation/ionization potential. These elements can be excited by collisions that take place at the interaction of the jet with the ambient medium, and depending on the energy involved in the shock, different species can be excited. Thus, emission lines with high ionization potential (e.g., [OIII]) trace strong shocks, while emission lines with low excitation/ionization potential (e.g., H$\alpha$+[NII] and [SII]) are associated with less energetic shocks.

 We can see in Figure \ref{orx} that the soft X-rays detected by Chandra (red contours) trace emission associated with the brightest knots identified in the optical range by \citealt{HRR1998} (HH~81A, HH~80A, HH~80G, and HH~80C, in gray scale), who also report their tangential velocities: 370$\pm$17~km~s$^{-1}$ (HH~81A), 334$\pm$23~km~s$^{-1}$ (HH~80A), 511$\pm$81~km~s$^{-1}$ (HH~80G), and 74$\pm$46~km~s$^{-1}$ (HH~80C). Faster ones (HH~81A, HH~80A, and HH~80G) present [OIII] emission, and also match radio continuum sources (blue contours), whose velocities are 351$\pm$104~km~s$^{-1}$ (HH~80) and 223$\pm$85~km~s$^{-1}$ (HH~81) (\citealt{masque2015}), in agreement with those estimated by \cite{HRR1998}. In the observations reported by \cite{masque2015}, knots HH~80A and HH~80G cannot be spatially resolved (beam=6''$\times$4''), and therefore, the HH~80 velocity represents the HH~80A-HH~80G system velocity. Considering a jet inclination of 49$^{\circ}$ (\citealt{girart2018}), the absolute velocities of these knots are: 564$\pm$26~km~s$^{-1}$ (HH~81A), 509$\pm$35~km~s$^{-1}$ (HH~80A), 779$\pm$123~km~s$^{-1}$ (HH~80G), 113$\pm$70~km~s$^{-1}$ (HH~80C), 535$\pm$158~km~s$^{-1}$ (HH~80), and 340$\pm$130~km~s$^{-1}$ (HH~81). On the other hand, black contours clearly reveal that hard X-ray emission arises from a very compact source, which is not associated with any optical or radio structures tracing shocks.

 The high angular resolution images in Figure \ref{orx} allow us to identify the regions emitting radio and soft X-ray radiation, associated with HH objects. We can see that soft X-rays emission present a morphology elongated in the direction perpendicular to the jet axis, and is located in front of the radio emission peak of HH ~81A, HH~80A, and HH~80G. The separation of radio and X-ray emission peaks in the plane of the sky are about 3000 au ($\sim 1\farcs8\pm0\farcs5$) in the case of HH~81A, and 1000 au ($\sim 0\farcs6\pm0\farcs5$) in the case of HH~80A and HH~80G. Here, errors were calculated as the quadratic sum of the astrometric accuracy in the X-ray (0.5 arcsec) and radio (0.1 arcsec) images. Therefore the observed offset seem to be real. On the other hand, the spectral analysis indicates that soft X-rays are tracing thermal emission produced by high temperature gas ($\sim 10^{6}$~K), while centimeter emission is dominated by optically thin synchrotron radiation, due to its negative spectral indices ($\alpha=-0.3\pm0.1$; \citealt{marti1993}).

 In the context of a jet interacting with the parent cloud, these results indicate that the observed emission can be naturally explained by the formation of a radiative forward shock, and an adiabatic reverse shock (Mach disk). In a shock that propagates with velocities of several hundreds km~s$^{-1}$ (as is the case of HH 81~A, HH~80A, and HH~80G), material shocked in the molecular cloud can be heated by collisions, reaching high temperatures (e.g., $\sim 10^{6}$~K, derived from the spectral fit). Gas at such temperatures produce X-ray thermal emission. On the other hand, non-thermal emission detected at radio frequencies can be interpreted as synchrotron radiation produced by particles accelerated in the reverse shock. Thus, initially thermal particles, could increase their energy to relativistic values by diffusing back and forth across the Mach disk, via DSA (e.g., \citealt{krymskii1977}, \citealt{axford1977}, \citealt{bell1978a}a, \citealt{bell1978b}b, \citealt{blandford1978}). This scenario is consistent with results obtained in shocks observed in the triple radio continuum source in Serpens (\citealt{arrk2016}) and in the inner region of the  HH~80-81 jet (\citealt{arrk2017}).

 Regarding the case of HH~80C, we can see that this object does not present emission detected in radio nor [OIII]. Its soft X-ray emission is associated with optical emission corresponding to low ionization/excitation potential species (H$\alpha$+[NII], [SII]), indicating that this is the case of a relatively weak shock, in agreement with its velocity, considerably lower (i.e., 74$\pm$46~km~s$^{-1}$). In this case, the jet-cloud interaction could increase the temperature of the shocked gas and produce thermal emission detectable at soft X-rays, but it would not be energetic enough to accelerate particles at relativistic energies in the Mach disk and produce non-thermal emission at radio frequencies.

\section{Jet termination shocks}
\label{termination-shocks}

\begin{figure*}
   \includegraphics[scale=0.4]{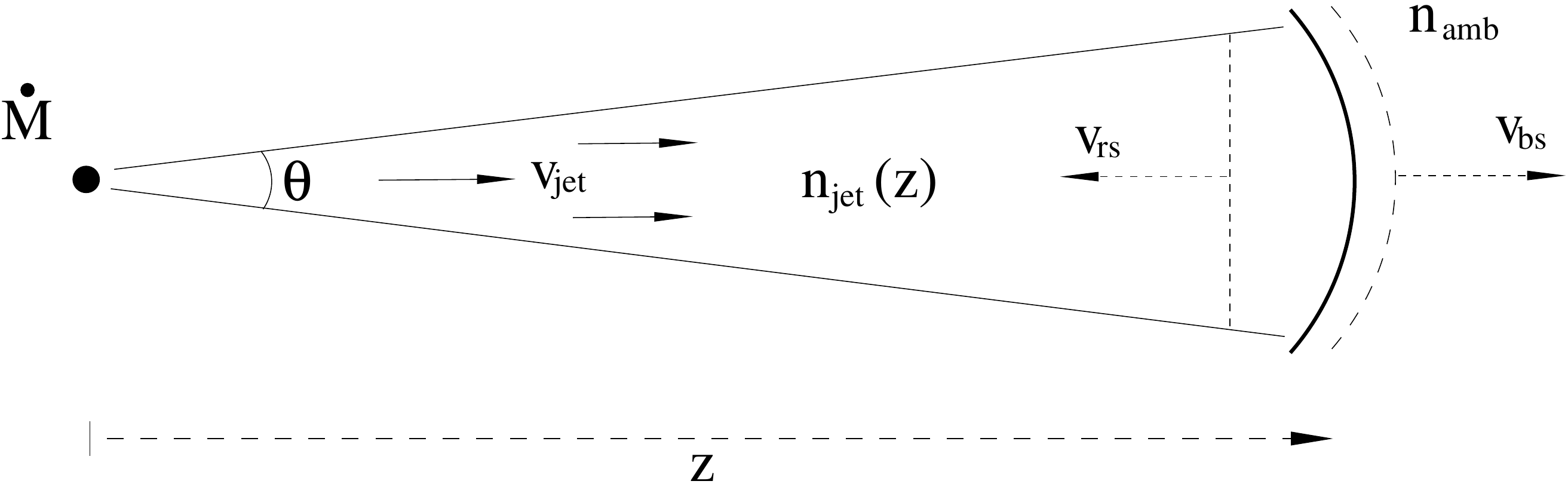}
   \caption{Scheme: the leading working surface at the jet termination region is 
composed by a bow shock and a reverse shock. Here $\dot{\rm M}$ represents the mass-loss rate, $\theta$ the opening angle of the flow, and $\rm v_{rs}$, $\rm v_{bs}$, $\rm n_{jet}$ and $\rm n_{amb}$ the reverse and forward shock velocities, and jet and ambient number densities, respectively. Figure extracted from \citet{arrk2016}.}
   \label{shock}
   \end{figure*}

The leading working surface at the jet termination region is 
composed by a bow shock and a reverse shock (or Mach disc), as is shown in
Figure \ref{shock}.
The multiwavelength data analyzed in this paper indicate that
X rays come from the bow shock in the external medium, whereas synchrotron
radio emission is produced in the jet reverse shock.

\subsection{Thermal emission from the bow shock}
\label{thermal}

The X-ray emission from HH~81A and HH~80A is consistent with thermal emission from diffuse gas at a temperature of $\sim 10^{6}$~K, implying a 
shock velocity $\sim 300$~km~s$^{-1}$, in agreement  with proper motion
measurements. The thermal cooling length d$_{\rm th}$ in the shocked plasma behind the bow shock is
\begin{equation}
\label{d_th}
\left(\frac{d_{\rm th}}{\rm cm}\right) \sim 3\times10^{16}\left(\frac{n_{\rm amb}}{100\,\rm cm^{-3}}\right)^{-1} 
\left(\frac{\rm v_{\rm bs}}{300\,\rm km\,s^{-1}}\right)^{4.5}
\end{equation}
(e.g. \citealt{raga2002}), being n$_{\rm amb}$ the density of the medium in which the jets propagates, and v$_{\rm bs}$ the bow shock velocity. We can see from Figure~\ref{orx} that 
the extension of the soft X-ray emission (red contours) is $\sim 2$~arcsec ($\sim 5\times10^{16}$~cm), and therefore the
density of the molecular cloud has to be $n_{\rm mc}> 100$~cm$^{-3}$ for the bow shock to be radiative. In the opposite case, i.e. $n_{\rm mc} < 100$~cm$^{-3}$, the bow shock would be adiabatic.
Radiative bow shocks are expected to be thin regions given the condition 
$d_{\rm th} < r_{\rm j}$ (e.g. \citealt{blondin1990}), whereby $r_{\rm j}$ is the jet cross-sectional radius at the position of the bow shock (i.e. HH~80) from the protostar. Here, the derived charateristic dimension of the thermal emitter ($\sim 5\times10^{16}$~cm) can be also the result of the poor spatial resolution of the instrument; therefore we are not able to specify if the bow shock is adiabatic or radiative.

However, HH~80 and HH~81 are located in the outskirts of the molecular cloud
where it is expected to be a diluted environment. Then, even when the bow shock is adiabatic, shocks with 300~km~s$^{-1}$ in a partially-ionized plasma (as molecular clouds) are not expected to be good particle accelerators given that diffusive Alfven waves are damped by ion-neutral collisions (e.g. \citealt{drury1996}, \citealt{padovani2015}).

\subsection{Synchrotron radio emission from the reverse shock}

According to the scenario proposed in this study, the non-thermal emission at 5~GHz can be produced by
a population of relativistic electrons accelerated via the Fermi~I
mechanism in the jet reverse shock (\citealt{bell1978a}). We consider that non-thermal electrons in HH~80A and HH~81A follow a power-law energy distribution with the $-$2 canonical index. In the equipartition regime, with non-thermal electrons, the magnetic field in the reverse shock downstream region can be estimated as:

\begin{equation}
\label{EqB}
\left(\frac{B_{\rm eq}}{\rm mG}\right)\sim 0.25
\left(1+a\right)^{\frac{2}{7}}
\left(\frac{f}{\rm mJy}\right)^{\frac{2}{7}}
\left(\frac{V}{4\times10^{48}\rm cm^{-3}}\right)^{-\frac{2}{7}}.
\end{equation}

\noindent where $a$ is the ratio of non-thermal protons to electrons, $f$ is the flux density at $5.5$ GHz, and $V$ is
the volume of the synchrotron emitter. 
From Figure \ref{structure} we can see that the synchrotron emission (at 5.5 GHz) in 
HH 80A presents a more circular geometry, whereas HH 81A is more elongated. Therefore, we assume that $V = (4/3)\pi r^3$ in the former case, and $V = (4/3)\pi r^2 l$ in the latter, being $l\sim7\times10^{16}$~cm and $r\sim5\times10^{16}$~cm dimensional parameters derived from Gaussian fits to the emission of the knots in our images. Thus, for an estimated flux density of $\sim$500 $\upmu$Jy, and considering $a = 0$ and $a = 40$ (\citealt{BK2005}), Equation \ref{EqB} gives a magnetic field of the order $B_{\rm eq}\sim0.1$~mG. It is worth noting that values of $B_{\rm eq}\sim0.1$~mG have been obtained in other protostellar jets such as IRAS~16547-4247 (\citealt{araudo2007}) and the central part of HH 80-81 (Carrasco-Gonz\'alez et al. 2010).

\section{Conclusions}

 We analyzed observations in the radio (VLA), optical (HST), and X-ray (Chandra and XMM-Newton) ranges of the Herbig-Haro objects HH~80 and HH~81.  The radio observations we present in this work are the highest angular resolution to date, allowing to compare with much more precision the emission regions and their morphology, with optical and X-ray radiation. Likewise, we studied the spectral nature of these objects at radio frequencies and X-rays. The object HH~80N, detected at centimeter wavelengths, reveals a clear bow shock structure, observed for the first time in a Herbig-Haro object in this spectral range. The results we obtained were interpreted in the context of shocks produced by the interaction of a protostellar jet with the surrounding medium, and can be summarized as:\\

\begin{enumerate}
\item High resolution radio images allow us to resolve the object HH~80 in two components (A and G, according to its identification at optical wavelengths). Both sources and HH~81 are elongated in the jet direction, and have its maximum emission in the apex, being consistent with what is expected in a bow shock formed by the interaction of a supersonic jet with the ambient medium. 

\item  The soft X-ray emission in HH~80 (A and G) and HH~81 is observed ahead of the radio peak and shows a structure elongated in the direction perpendicular to the jet axis.

\item HH~80N reveal a complex structure that was not previously detected in HH objects, i.e., flattened in the frontal region of the shock, with a slightly curved bow-like shape, and a faint tail in the jet direction (towards the protostar).\\

\item Objects HH~81A, HH~80A, and HH~80G identified in the optical range, also present radio continuum (5.5~GHz) and soft X-ray (0.3-1.2~keV) emission. These objects trace strong shocks associated with [OIII] emission, in agreement with their high velocities (300-500~km~s$^{-1}$). From the spectral analysis we see that the X-ray emission can be interpreted in terms of thermal radiation produced by hot diffuse gas at high temperatures ($\sim 10^{6}$~K). On the other hand, previous works have reported radio emission with negative spectral indices, indicating the existence of a non-thermal component in these objects.\\

\item From items (1), (2), and (4) it follows that the observed emission can be naturally explained by means of a radiative forward shock, and an adiabatic reverse shock (Mach disk), that take place in strong interactions of the jet with the ambient medium. In the radiative forward shock, material shocked in the molecular cloud can be heated by compression and reach temperatures high enough to produce soft X-ray thermal emission (ahead of the radio emission). On the other hand, an adiabatic reverse shock would be able to accelerate particles up to relativistic energies, producing synchrotron emission detectable at radio frequencies behind the forward shock.\\

\item HH~80C does not present radio nor [OIII] detected emission, and its emission at soft X-rays corresponds to optical emission of low ionization/excitation potential (H$\alpha$+[NII] and [SII]). This implies that it would be the case of a low velocity (74$\pm$46~km~s$^{-1}$) relatively weak shock, strong enough to rise the medium temperature and produce soft X-rays emission in the forward shock, but not energetic enough to accelerate particles in the Mach disk and produce detectable non-thermal radio emission.
\end{enumerate}

\noindent Acknowledgments: This work was supported by UNAM DGAPA-PAPIIT grant number IA102816, IN10818. We thank the anonymous referees for their comments which improved the manuscript. P.H. acknowledges partial support from NSF grant AST-1814011.

\begin{table*}
   \caption{Images Parameters. We present detailed information about the radio images we worked with. In the first column we list the year at which the data were obtained. Second column correspond to the central frequency observed with a bandwidth listed in column (4). The third column correspond to the VLA configuration, which refers to different antenna configuration. In column five we specify the weighting used to made the images. This allows to obtain different angular resolution, as can be seen in column (6). The synthesized beam correspond to a bidimensional Gaussian, whose major axis has a position angle listed in column (7).  \label{tbl-images_param}}
  \begin{center}
    \begin{tabular}{ c c c c c c c}
    \hline\hline
   Epoch & Frequency  & Configuration & Bandwidth  & Weighting & Synthesized    &   PA\\
   {}   &  (GHz)   &  {}   &    (GHz)   &  {}   &   Beam   &   {}\\
    \hline
   2012$^{a}$              &    5.5          &    B    &    2.0       &    Robust = 1.5    &    $2\farcs19\times 1\farcs32$   &   $4\fdg5$     \\
   2009$^{b}$              &    5.0          &    C    &    0.1       &    Natural         &    13$''$ $\times$ 8$''$   &   2$^{\circ}$       \\
   2009/2012$^{c}$         &    5.5          &    C/B  &    2.0       &    Robust = 0      &    $2\farcs92\times 1\farcs83$   &   $2\fdg4$     \\   
   2009/ 2012$^{c}$        &    5.5          &    C/B  &    2.0       &    Robust = -1     &    $1\farcs64\times 0\farcs91$   &   $2\fdg8$     \\
  \hline
  \end{tabular}\\
  \end{center}
   \small{$^{a}$ Reported by \citealt{arrk2017}\\
   \small{$^{b}$ Reported by \citealt{carrasco2010Sci}.}\\
   \small{$^{c}$ Combined data from $^{a}$ and $^{b}$.}
   }
\end{table*}


 \begin{table*}
  \caption{Spectral fit parameters. We present the parameters obtained from a spectral fit to soft X-ray emission associated with the HH objects (listed in the first column). Columns (2), (3), and (4), are derived from a thermal fit and correspond to the hydrogen column density, visual extinction, and energy, respectively. The energy allows to calculate the temperature of the emitting gas (column 5), and the shock velocity (column 6). In column (7) we list the photon index $\upgamma$ of the power law fit. \label{tbl-spec}}
  \begin{center}
    \begin{tabular}{ c c c c c c c }
    \hline\hline
   Object & $N_{H}$               & $A_{V}$             & kT                     &  T                  & $V_{\rm shock}$  & $\upgamma$ \\
   HH     & (10$^{22}$~cm$^{-2}$) & mag                 & (keV)         & (10$^{6}$~K)        & (km~s$^{-1}$)     & {} \\
   \hline
   HH~81  & 0.7$_{-0.3}^{+0.2}$   & 3$\pm$1             & 0.10$_{-0.10}^{+0.05}$ & 1.3$_{-1.2}^{+0.6}$ & {290}$_{-136}^{+68}$             & 0.9$\pm 0.2$\\
   \\
   HH~80  & 0.6$_{-0.6}^{+0.2}$   & 2.8$_{-2.8}^{+0.8}$ & 0.11$_{-0.03}^{+0.05}$ & 1.3$_{-0.3}^{+0.6}$ & {293}$_{-34}^{+68}$             & 0.9$\pm 0.2$\\
  \hline
  \end{tabular}\\
  \end{center}
  \small{Errors with 1$\sigma$=68\% confidence.}
\end{table*}




\bsp	
\label{lastpage}
\end{document}